# Understanding the Skills Gap between Higher Education Institutions and the Software Engineering Industry

Huy Phan[1], Ievgeniia Kuzminykh[1,*], and Bogdan Ghita[2]

[1]Department of Informatics, King's College London, Strand, WC2R 2LS, UK
[2]School of Engineering, Computing and Mathematics, University of Plymouth, PL4 8AA, UK
filipphan29@gmail.com, ievgeniia.kuzminykh@kcl.ac.uk, bogdan.ghita@plymouth.ac.uk

**Abstract**

In the rapidly evolving field of software engineering, the skills required of graduates entering the job market are constantly changing. Several studies have identified a gap between the skills taught in university curricula and those demanded by the software engineering industry. This chapter investigates the technical skill and expertise gap between higher education institutions (HEIs) and the UK software engineering industry by mapping job descriptions to the skills included in computer science degree programmes. A custom web scraping and text analysis tool, utilising fuzzy matching, was developed to extract and categorise skills from 300 job postings and undergraduate curricula from 30 UK universities. The analysis showed that the curricula place a strong emphasis on Programming Languages (18%) and Database Management (12.83%). In contrast, the industry's most frequently requested skill category is Software Design and Planning, which appears in approximately 88.68% of job descriptions, highlighting its critical importance. General Programming Language and System Structures also show strong demand, present in over 78.30% and 66.04% of postings, respectively. The mapping indicates that areas such as System Structures and Software Domains are significantly underrepresented in curricula, while Database Management and Compiler Design may be overemphasised. These insights can support HEIs in aligning their programmes with industry needs, supporting the preparation of graduates for dynamic careers in software engineering.

## 1 Introduction

The skills required of IT and software engineering graduates are evolving rapidly in response to technological advancements [1]. Numerous studies and industry reports indicate that computer science graduates often enter the workforce without fully meeting the demands of the technology sector, revealing a persistent gap between higher education curricula and industry needs [2, 3]. To identify specific skill gaps, a range of approaches has been proposed, including surveys, the aggregation of quantitative data from existing studies, and the application of probabilistic topic modelling techniques [2, 4, 5]. It is also important to recognise that addressing the gap remains a complex and ongoing challenge, partly due to the difficulties higher education institutions (HEIs) face in reviewing and updating curricula. [6]. Therefore, HEIs must remain aware of the skill gaps, and proactively recognising and mitigating these gaps is essential, not only to better prepare graduates for the workforce but also to prevent broader economic consequences [7].

In this chapter, we aim to explore the technical skill and expertise gaps between higher education and the software engineering (SE) industry in the United Kingdom. To do this, we categorise the technical skills required in job advertisements and those offered in computer science degree programmes, and then systematically map them. Ultimately, this analysis is intended to answer the central research question:

*What are the prominent technological knowledge gaps between HEI and the SE industry in the UK?*

To achieve this aim, the study in this chapter will pursue the following objectives:
- Identifying the skills demanded by industry through the analysis of job advertisements.
- Determining the skills and expertise offered by UK university computer science programmes.
- Mapping the frequency of skill categories across HEIs and industry.
- Highlighting the skill gaps between HEIs and industry.
- Visualising the results and key findings.

The research involves the development of a web scraping tool and a text analysis tool to automatically extract and classify skills from job descriptions. The core analytical approach method involves comparing the frequency of different skill categories between job postings and curricula. The findings of this study aim to inform the design of future computer science programmes, ensuring that graduates are better prepared for successful careers in software engineering.

The remainder of this chapter is organised as follows. Section 2 reviews existing studies to establish a solid background. Section 3 outlines the requirements for the web scraping and text analysis tool, describing the methodology employed in the study. Section 4 presents the analysis of the results for the skill gap and a critical evaluation of the study's outcomes. Section 5 discusses the obtained results and compares them with existing studies, concluding with suggestions for future directions in Section 6.

## 2   Related Works

The literature demonstrates that aligning software engineering curricula with industry requirements remains a key research challenge. This section reviews a wide range of methodological approaches adopted by researchers, including systematic literature reviews, thematic analysis of questionnaires and interviews, and the application of job advertisement scraping alongside probabilistic topic modelling techniques. While this review provides valuable insights into the diversity of methods used to investigate skill gaps, it also highlights inherent limitations, suggesting that further methodological refinement and complementary approaches are needed to more effectively investigate and address skill gaps.

A study by Garousi et al. [4] sought to identify areas of software engineering that exhibit the most substantial knowledge shortfalls by conducting a literature review of 33 studies addressing the alignment of SE education with industry expectations. Quantitative findings from the selected primary studies were synthesised to present an aggregated perspective on the topic. The results revealed significant knowledge gaps in areas including design, testing, and configuration management. A threat to validity arises from the possibility that the final set of 33 papers may not comprehensively reflect the full scope of research, given the reduction from an initial pool of 94 studies. Moreover, the focus on publications from 12 countries, such as the USA, South Africa, and Spain, spanning the period from 1995 to 2018 may overlook more recent developments and emerging trends in both academia and industry. In contrast, the study in [8] reviewed literature published between 2011 and 2020, concentrating on the reasons of skills gaps and corresponding mitigation strategies rather than identifying the gaps themselves.

In another notable study by Oguz et al. [2], interviews, questionnaires, and surveys involving students and recent graduates were combined to uncover the principal difficulties faced when entering professional practice. The results indicated that requirements analysis and software design pose particular challenges for students. Furthermore, a quantitative and qualitative analysis of 50 job advertisements identified strong employer demand for database and programming skills. A key strength of this study is that it incorporates perspectives from academics, students, and recent graduates. However, the findings are limited by their breadth, as the category of 'programming' encompasses a broad range of skills. A more focused analysis of specific programming languages in demand could yield more actionable insights. Similar mixed-method approaches using surveys and interviews were reported in other studies [1] [6], which identified gaps in areas such as agile software development, quality assurance practices, and testing.

The study by Gurcan et al. [5] demonstrates the potential of probabilistic topic modelling to identify dominant technologies and emerging skill combinations within the software engineering job market. By applying Latent Dirichlet Allocation (LDA) to job advertisements collected from Stack Overflow Careers, the study showed how large-scale, real-world data can be used to identify dominant technologies and frequently mentioned skill sets. LDA discovers abstract topics across a collection of documents by modelling each document as a combination of topics, and each topic is defined by a probabilistic distribution of words [9]. The findings highlight the increasing importance of combined front-end development skills, such as HTML, CSS, and JavaScript, which were identified as the most in-demand skill set, at 12.4%. Meanwhile, Java appeared most frequently overall, featuring in 21% of job postings. However, a limitation of this method lies in the challenge of selecting suitable LDA parameters, particularly the number of topics, a process that is inherently subjective and requires multiple experimental iterations.

The study in [10] conducts a systematic mapping of trends in software engineering to explore the relationship between industry practice and SE education. Relevant publications were retrieved from multiple databases and subsequently filtered based on predefined criteria. The selected papers were then keyworded and organised into a classification framework encompassing prevalent SE trends. The results showed that Agile Software Development is the most common trend, while Systems of Systems (including cloud systems, mobile systems, etc.) is the least common. The use of classification makes it easier to identify patterns and gaps in the research landscape, which is especially effective in fields like software engineering, where the body of research is large and diverse. The study [11] also utilised systematic mapping, which revealed a trend of tech start-ups adopting lean methodology and global software engineering being more prominent in SE education.

Regarding the collection of data from job advertisements, the study by Hanna [12] provides a large-scale comparison between software testing education and industry requirements. Through the analysis of 80 testing-related courses across nine countries and 400 job advertisements, the authors identify a clear mismatch between academic coverage and the skills most frequently demanded by employers. The results showed a clear mismatch with many industry-relevant skills being either minimally covered or entirely absent from the curricula. In particular, topics such as web and mobile application testing, requirements traceability, troubleshooting, agile and DevOps, and the use of automation and testing tools frequently appear in job advertisements but are rarely covered in course content. The analysis also highlights wide variation in testing topics and course emphasis across universities, even within the same country, suggesting a lack of alignment in educational approaches. The study by Hamid and Ikram [13] is based on Hanna's study, but it analysed only the job advertisements' requirements for software testers and was limited to Pakistan. Both studies are focused only on software testing, while we aim to cover the entire SE curriculum.

A practical approach to job market analysis was proposed by students at Ostured University through the development of the Job Market AnalyseR (JMAR), a tool designed to automatically import data and conduct text analysis [3]. The process begins with data import, followed by a procedure which generates a table showing skill frequencies based on terminology derived from Stack Overflow. This is accomplished by scanning job advertisement text for occurrences of keywords outlined in a predefined skills table. A skill ratio is then calculated across the full document set. The analysis revealed a growing emphasis on cloud computing and automation technologies, such as Docker and Kubernetes, which were largely absent from the reviewed curricula, highlighting a tangible misalignment between educational offerings and industry needs. One limitation of this study is that it focuses on specific programming languages and technologies, without addressing higher-level competencies such as software testing or design.

In summary, the methodologies employed to detect gaps between HEI and industry display several shortcomings. Some studies adopt overly general descriptors, such as the term "programming", whereas others are narrowly focused on specific programming languages. Additionally, approaches that rely on previously published literature are unable to capture the dynamics of the current job market and may therefore overlook newly emerging technologies. Moreover, none of the reviewed studies explicitly targets the UK job market. A summary is also provided in Table 1. In response, the development of job-scraping and text-analysis tools appears well-suited to support automated and scalable analysis of labour market trends, particularly as technologies continue to evolve. Categorising skills can also facilitate a higher-level perspective, enabling clearer identification of existing skill gaps, given that academic modules typically address broader areas of knowledge rather than specific skills.

Table 1. The studies investigating the gap between HEI and industry demands in software engineering

| Study | Year | Location | Method of Analysis | Identified Skill Gaps |
|---|---|---|---|---|
| [4] | 2020 | Global | Literature Review of 33 Studies from 1995-2018 | Configuration management, design, testing |
| [2] | 2019 | Turkey | Interviews, Questionnaires, and Surveys | Software design, requirements analysis, programming and database skills |
| [1] | 2019 | Global | Surveys | Agile software development and testing |
| [5] | 2016 | Global | Latent Dirichlet Allocation (LDA) Model | Combinations of programming languages (HTML, CSS, JavaScript), Java |
| [9] | 2021 | Global | Systematic Mapping | Agile software development, systems of systems |
| [3] | 2023 | Sweden | Job Market AnalyseR (JMAR), Text Analysis | Cloud and automation technologies (Kubernetes, Docker) |
| [8] | 2024 | Global | Literature review of 31 studies from 2001-2020 | Not specified, focus of the paper on causes of skill gap and mitigation strategies |
| [12] | 2022 | Global | Focus on software testing courses, 80 course and 400 job adverts Mapping | Testing in agile development, DevOps, Unit testing, System testing and many others |
| [13] | 2024 | Pakistan | Analysis of 160 job advertisements | Testing skills with emphasis on testing process and management. |

# 3 Methodology

This section outlines the methodology adopted in this chapter, and a high-level diagram of this methodology can be seen in Figure 1. The methodology involves several phases: collecting job advertisements and curricula data, extracting skill categories, and analysing the data. We followed the approach presented in the studies [14] [15], where each phase systematically investigates the alignment between educational offerings and market needs.

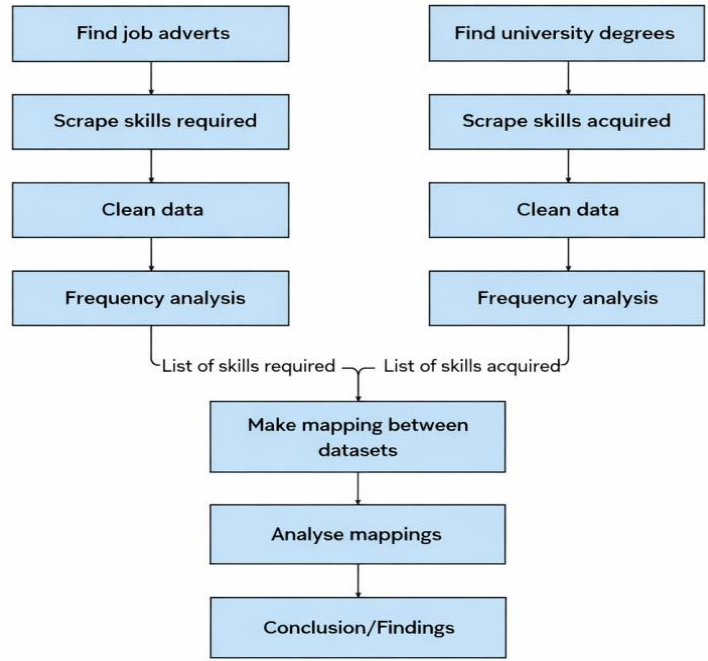

Figure 1. Methodology for collecting and analysing the curricula and job postings

## 3.1 Data Collection

To assess industry skill requirements, job advertisements were systematically collected using a custom web scraper developed for Indeed.com. Indeed was selected due to its standardised HTML structure, which facilitates consistent data extraction across listings, as well as its widespread use in related literature [16] [17] [18] [19] and its large volume of postings [20]. The scraper, built using Python and Selenium, effectively navigates and extracts information from dynamic web pages. The tool focuses on capturing key details such as job title, location, salary, and description, which are then stored in a CSV file.

The tool was set up to collect job advertisements according to defined parameters: the geographical location was the United Kingdom, the employment type was limited to full-time roles, the required education level was a bachelor's degree, and the role was software engineer. These selection criteria ensure that the collected job descriptions closely align with the study's focus on UK undergraduate curricula within HEIs. For instance, full-time roles were prioritised because such positions typically represent the main employment pathway for graduates seeking employment after completing higher education [21].

On the academic side, course-related data from leading UK universities was collected using a manual collection process. This approach was adopted because each university website has a different structure, making the creation of a universal web scraper impractical. Universities were selected according to their rankings in the Complete University Guide [22], with particular attention given to computer science and SE programmes, as graduates from these disciplines commonly apply for the same positions. Furthermore, excluding Computer Engineering programmes that offer courses in electronics and hardware would be inappropriate. The data collected encompass curriculum information, including course titles, university, and the modules delivered within each programme. While more detailed curriculum specifications are available in accreditation reports, these documents are frequently not accessible to the public.

### 3.2 Data Processing

Since both datasets are unstructured, we performed data cleansing by removing stop words, generating n-grams (sequences of n consecutive words), and counting the number of occurrences for each skill. The use of the Natural Language Toolkit (NLTK) [23] enables the automation of these processes, improving efficiency and scalability compared to manual analysis. Normalisation steps, including lowercasing, tokenisation and the removal of punctuation and non-alphanumeric symbols, further enhance data consistency and reduce noise, thereby supporting more reliable downstream analysis.

### 3.3 Data Analysis

The analysis phase included an exploratory analysis of the job adverts and university courses separately. Skills extracted from job advertisements and modules were categorised and mapped accordingly, which requires defining skill categories. The definition was guided by two software engineering taxonomies:
- ACM Computing Classification System (ACM CCS) [24]
- SoftwareEngineering Body of Knowledge (SWEBOK) [25].

The ACM CCS taxonomy uses a multilevel hierarchy list of keywords to organise topics, as shown in Table 2 for L1 and L2 layers.

Table 2. Example of L1 and L2 categories of ACM CCS

| L1 | L2 |
|---|---|
| Software organisation and properties | Contextual software domains |
| | Software system structures |
| | Software functional properties |
| | Extra - functional properties |

The SWEBOK taxonomy outlines universally accepted knowledge areas (KAs) in software engineering, dividing the field into 15 KAs, such as Software Requirements, Software Design, Software Construction, and Software Testing. The use of established taxonomies for categorisation and mapping has been applied successfully in prior studies [16] [17] within the field of cybersecurity rather than software engineering; nevertheless, the underlying process is comparable.

Categories that did not directly relate to technical skills were excluded, such as collaboration-focused areas in the ACM taxonomy and software engineering economics in the SWEBOK. Related categories were then merged to reduce overlap; for example, compiler-related topics were combined into a single programming language theory category. This process resulted in ten skill categories presented in Table 3, which were further grouped into three broader areas aligned with the higher-level structure of the ACM Computing Classification System.

Modules can be placed into multiple categories as they address multiple knowledge areas. For example, the Embedded Systems Programming module could be categorised under System Structures and General Programming Languages.

Table 3. Skill categories extracted from SWEBOK and ACM CCS taxonomies for software engineering

| Software organisation & properties | Software & its engineering | Software creation & management |
|---|---|---|
| Software Domains | General Programming Languages | Software Design and Planning |
| System Structures | Database Management | Software Development Techniques |
| | Programming Language Theory and Compiler Design | Software Verification and Validation |
| | Development Frameworks and Tools | |
| | Configuration Management and Version Control | |

As fuzzy keyword matching is applied to the analysis of job advertisements, each skill category must be supported by an associated set of keywords. The selected keywords were derived from the lower levels of the ACM CCS taxonomy and from the descriptions of knowledge areas in SWEBOK. Where specific technologies were included as keywords, those with the highest levels of popularity were chosen. The complete set of skill categories and their corresponding keywords is provided in Table 4. It is important to note that, following the execution of fuzzy matching, single-letter technologies such as *C* and *R*, as well as terms with multiple meanings, including the programming language *Go*, were removed from the keyword list due to the high number of false positives. A similar issue was reported by Felix et al. [3], whose fuzzy-matching approach likewise required the exclusion of single-letter technologies.

Table 4. Skill categories in software engineering and associated keywords

| Category | Abbreviation | Keywords |
| --- | --- | --- |
| Software Domains | DOM | E-Commerce, virtualisation, OS, fintech, banking, healthcare, e-learning |
| System Structures | SYS | architecture, embedded, distributed, real-time, large-scale, web, microservices, cloud, modularity |
| General Programming Languages | PROG | Python, JavaScript, Java, C#, C++, PHP, HTML, CSS, TypeScript, PowerShell, Kotlin, Bash, Ruby, Rust |
| Database Management | DATA | database, SQL, MySQL, NoSQL, PostgreSQL, SQLite, MongoDB, RDBMS, Couchbase, Cassandra |
| Programming Language Theory and Compiler Design | PLT | formal, compiler, parsers, generation, syntax, lexical, semantics, correctness, Theory and Compiler Design interpreter |
| Development Frameworks and Tools | FRWK | Angular, ASP.NET, React, Django, Express, Spring-Boot, Flutter, Laravel, jQuery, Xamarin |
| Configuration Management and Version Control | CONF | Ansible, Git, GitLab, GitHub, SVN, Mercurial, Perforce, Kubernetes, Docker |
| Software Design and Planning | DES | design, specification, requirements, planning, implementation, UML, prototyping, modelling |
| Software Development Techniques | DEV | automation, OOP, TDD, CI/CD, flowcharts, risk, scrum, agile, DevOps, refactoring |
| Software Verification and Validation | VER | validation, verification, testing |

### 3.3.1 Frequency Analysis

For the analysis of job descriptions, the method employed fuzzy matching, a technique used to detect approximate or non-exact matches between textual elements [26]. This approach is particularly useful when equivalent technical skills are phrased differently in job advertisements, for example, "full stack development" and "end-to-end development". Latent Dirichlet Allocation, as described in [9] and applied in [5], although effective for identifying broader themes and topics, often lacks the ability to represent fine-grained contextual differences in keywords within a given document and to accurately identify specific skills [27].

For curriculum analysis, the method used fuzzy matching as well. In addition, the proportion of modules that deliver or are related to software engineering concepts was assessed. The curricula of the top 15-ranked and bottom 15-ranked university courses were then compared to determine whether there were any significant differences in curricula.

A similarity threshold of 95% was used for job description matching to classify a result as valid. For the grouping of module names, a lower threshold of 75% or above was applied.

### 3.3.2 Mapping Analysis

After completing the exploratory analysis of each dataset, the subsequent step involves mapping the frequencies of the identified from both higher education curricula and industry job postings. By quantifying and comparing skill categories across datasets, the analysis allows for a systematic identification of mismatches between what universities teach and what the industry demands. The identification of such gaps can inform the future development of university computer science programmes, support industry in accessing graduates with appropriate competencies, and contribute to ongoing research in this area. Mapping should be repeated periodically due to evolving industry needs and technological advancements [1].

## 4 Results and Analysis

### 4.1 University Curriculum Analysis

The final number of university curricula was 30, collected from undergraduate programmes offered by UK universities. It was compiled in November 2023 using the 2023 Computer Science subject ranking and course information provided by The Complete University Guide [22].

#### 4.1.1 Modules Analysis

The most common modules were extracted from curricula and grouped because modules may have slightly different names but the same content. The results are shown in Figure 2. The module *Software Engineering Design* ranks highest in frequency, with *Mathematics for Computer Science* following closely behind, indicating a strong academic emphasis on these subjects. *Artificial Intelligence* and *Object-Oriented Programming* are also well-represented, suggesting that these areas constitute common components of the curriculum.

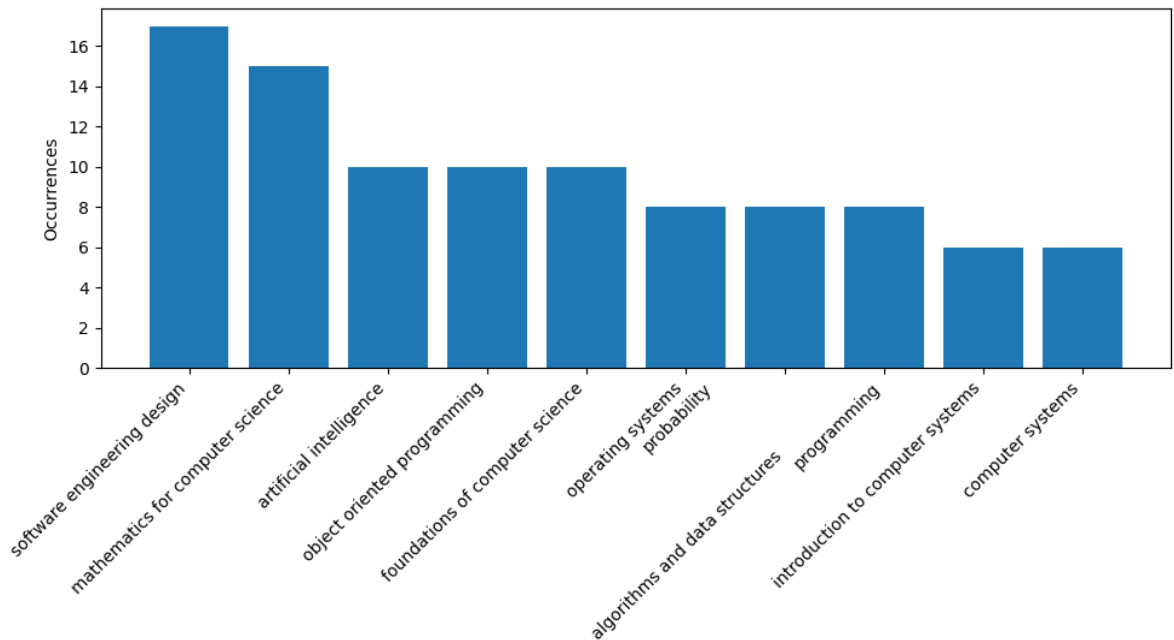

Figure 2. Most common modules in the curricula of the UK universities in undergraduate programmes in SE

#### 4.1.2 Curriculum Analysis by Skill Category

Table 5 shows the distribution of modules across different skill categories defined in Table 3 within the analysed Computer Science curricula. It should be noted that the percentages reported in Table 5 do not sum to 100%. This is because individual modules were allowed to be mapped to multiple skill categories when they addressed more than one knowledge area. For example, a module covering embedded systems programming may contribute to both the *System Structures* and *General Programming Languages* categories. As a result, the percentages represent the relative frequency of each skill category across the curriculum rather than mutually exclusive portions of the total module count.

Firstly, the curriculum places a strong emphasis on *Programming Languages* (18%) and *Database Management* (12.83%). These areas are fundamental to software engineering and suggest robust training in the technical skills necessary for many entry-level positions in the industry. *Software Design and Planning* (16.46%) receives a significant share of the curriculum, which is encouraging since it suggests a focus on the architectural aspects of software development. However, in light of the industry's increasing adoption of agile methodologies and rapid development cycles, there may be a need to place even greater emphasis on these areas in order to better equip students for the dynamic workflows they are likely to encounter [10].

The results show that several categories have comparatively low representation, including *Software Domains* (3.39%) and *Programming Language Theory and Compiler Design* (5.62%), which may suggest areas in which recent graduates are less well prepared. Industry frequently demands in-depth knowledge of particular application domains, such as finance or healthcare, while a sophisticated understanding of programming language theory can be important for certain advanced roles [6]. Reduced curricula emphasis in these areas may create a gap where graduates do not fully satisfy the specialised expectations of some sectors. Additional categories with limited coverage include *Development Frameworks* (7.9%) and *Configuration Management* (7%).

Table 5. Curriculum distribution by skill category

| Category | Percentage of total modules |
|---|---|
| Software Engineering modules (out of all curricula) | 37.70% |
| General Programming Languages | 18.00% |
| Software Design and Planning | 16.46% |
| Database Management | 12.825% |
| Software Development | 12.35% |
| System Structures | 8.23% |
| Software Verification | 8.24% |
| Development Frameworks | 7.89% |
| Configuration Management | 7.00% |
| Programming Language Theory and Compiler Design | 5.615% |
| Software Domains | 3.385% |

In summary, the curriculum analysis indicates comprehensive coverage of core software engineering skills, with 37.70% of modules allocated to this field. This number shows the overall proportion of curriculum content that can be classified as software engineering–related, aggregated across all defined skill categories. The strong emphasis on programming languages, software design, and planning implies that students are generally well prepared in these fundamental domains. Nevertheless, the comparatively limited focus on areas such as *Software Domains* and *Programming Language Theory* remains noteworthy.

### 4.1.3 Comparing Top 15 and Bottom 15 Universities

Table 6 compares the curriculum focus between the top 15 universities and the bottom 15 universities, as ranked by the Complete University Guide [22]. Software Engineering modules collectively account for a larger portion of the curriculum in the top 15 universities (43.81%) compared to the bottom 15 (39.91%). It represents the percentage of modules within a degree programme that address at least one software engineering skill category (as defined using ACM CCS [24] and SWEBOK [25]). Because modules can be mapped to multiple skill categories, this total is not a simple sum of the individual category percentages.

This suggests that higher-ranked institutions may place greater emphasis on software engineering as a distinct and substantial component of computer science education. Differences are particularly visible in areas such as *Programming Language Theory and Compiler Design* and *Database Management*, which receive more attention in the top 15 universities. In contrast, *Software Development* techniques are more prominent in the bottom 15 universities, accounting for 14.68% of modules compared to 10.18% in the top group.

Beyond the categories already discussed, the distribution across most other skill areas is relatively similar, indicating a broadly consistent curriculum structure across institutions regardless of ranking.

Table 6. Curriculum focus of top 15 and bottom 15 universities

| Category | Top 15 Universities (%) | Bottom 15 Universities (%) |
|---|---|---|
| Software Domains | 3.10 | 3.67 |
| System Structures | 8.29 | 8.17 |
| General Programming Languages | 18.58 | 17.43 |
| Database Management | 13.72 | 11.93 |
| Programming Language Theory and Compiler Design | 6.64 | 4.59 |
| Development Frameworks | 7.52 | 8.26 |
| Configuration Management | 6.87 | 7.13 |
| Software Design and Planning | 15.49 | 17.43 |
| Software Development | 10.12 | 14.58 |
| Software Verification | 7.89 | 8.59 |
| Software Engineering modules (out of all curricula) | 43.81 | 39.91 |

## 4.2 Job Adverts Analysis

A total of 300 job advertisements were analysed and collected between December 2023 and March 2024.

### 4.2.1 Programming Languages

Figure 3 presents the frequency with which programming languages are mentioned in job descriptions, providing insight into the programming skills most in demand by the industry. JavaScript is the most frequently cited language, appearing in approximately 30% of job advertisements, highlighting its central role in modern software development. HTML and CSS are also prominently mentioned, reflecting the sustained demand for front-end development skills. Python and Java similarly show high frequencies, indicating their broad applicability across software domains. These findings align with prior studies that report strong industry demand for these languages [3, 5]. In contrast, languages such as Kotlin, Bash, Ruby, PowerShell, and Rust appear less frequently. While this may suggest lower overall demand, it more likely reflects their use in specialised or niche contexts rather than a lack of relevance.

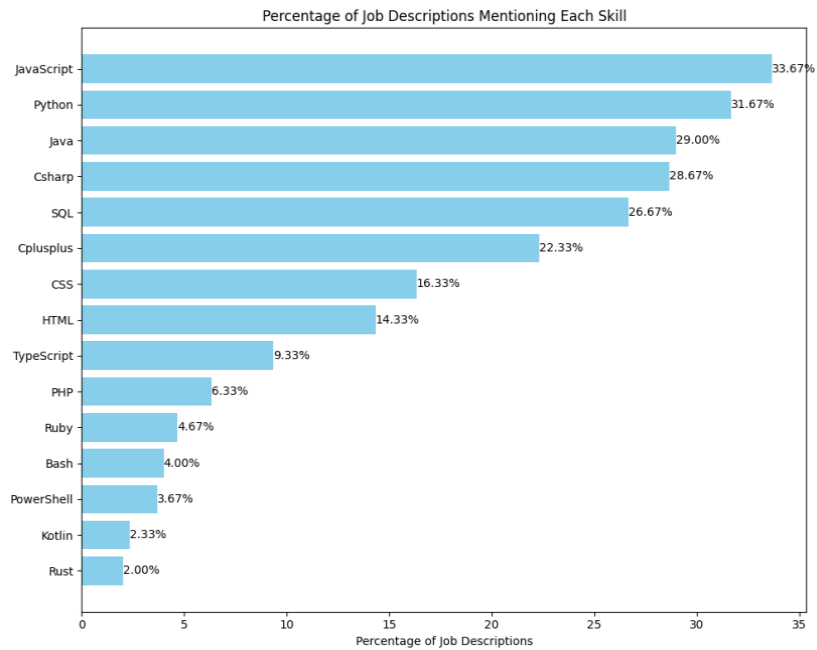

Figure 3. Frequency of programming languages in job adverts

### 4.2.2 Development Frameworks and Tools

Figure 4 illustrates the frequency of development frameworks and tools mentioned in job advertisements. React is the most frequently referenced framework, appearing in nearly 20% of job descriptions, signalling its dominance in contemporary front-end development. Angular and jQuery also feature prominently, indicating continued industry reliance on established frameworks alongside newer technologies. Overall, the results suggest that employers place strong emphasis on practical experience with widely adopted frameworks rather than exposure to a large number of tools.

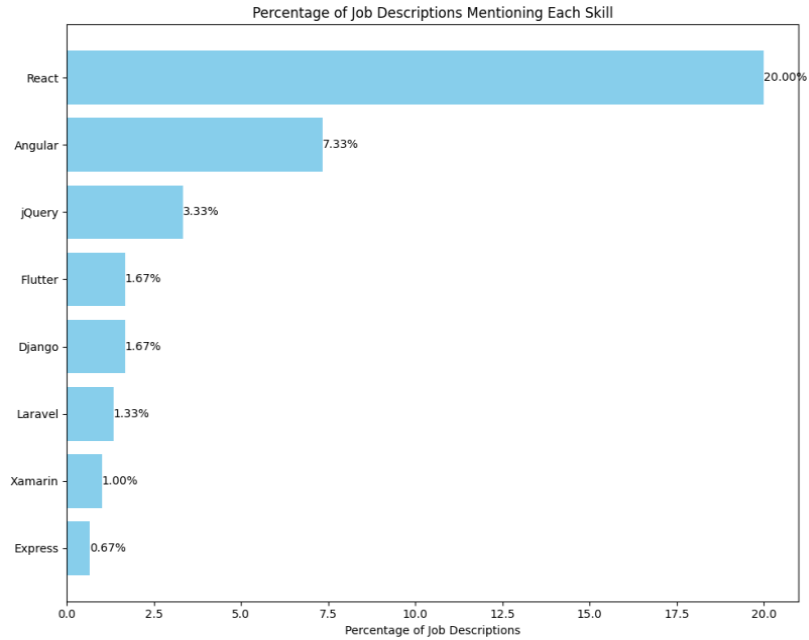

Figure 4. Frequency of development frameworks in job adverts

**4.2.3 Skill Categories in Job Descriptions**

Table 7 provides a consolidated overview of the demand for different software-related skills in the job market, offering key insights into potential gaps between higher education curricula and industry requirements.

*Software Design and Planning* emerges as the most frequently requested skill category, appearing in approximately 88.68% of job descriptions, underscoring its critical importance. *System Structures* and *General Programming Languages* also demonstrate strong demand, mentioned in 66.04% and 78.30% of job postings, respectively. In contrast, *Development Frameworks and Tools* and *Programming Language Theory and Compiler Design* appear less commonly, featuring in approximately 20.75% and 16.04% of advertisements, indicating that these skills are required for a smaller subset of roles. It is worth noting that each job posting can list multiple skills, so a single job description may be counted in several categories.

Table 7. Percentage and total frequency of skill categories in job advertisements

| Skill Category | % of Job Descriptions Mentioning Category | Total Skill Mentions |
|---|---|---|
| Software Design and Planning | **88.68%** | 547 |
| General Programming Languages | **78.30%** | 848 |
| System Structures | **66.04%** | 369 |
| Software Development Techniques | 46.23% | 213 |
| Software Domains | 45.28% | 114 |
| Software Verification and Validation | 40.57% | 167 |
| Configuration Management and Version Control | 25.47% | 142 |
| Database Management | 21.70% | 137 |
| Development Frameworks and Tools | 20.75% | 111 |
| Programming Language Theory and Compiler Design | 16.04% | 37 |

The total skill counts shown in Table 7 further enrich this analysis. Notably, *General Programming Languages* has the highest total number of skill mentions (848 occurrences), despite appearing in a smaller percentage of job descriptions than *Software Design and Planning*. This discrepancy can be explained by the fact that individual job advertisements frequently list multiple programming languages (e.g., JavaScript, Python, Java, SQL) within a single posting. As a result, while the category may appear in fewer job descriptions overall, it accumulates a higher total number of mentions due to repeated references within each advertisement.

In contrast, *Software Design and Planning* skills are typically referenced more broadly, resulting in fewer repeated mentions per job description. Similarly, *System Structures* and *Software Domains* display relatively lower total occurrence counts compared to the percentage of job descriptions in which they appear, suggesting that these skills are commonly required but are often described at a higher level of abstraction. This distinction highlights the importance of considering both metrics when assessing industry skill demand, as percentage coverage reflects breadth across roles, while total counts indicate depth of emphasis within individual job descriptions.

### 4.3 Mapping Analysis

To compare the skills provided by universities with those demanded by industry, the analysis considers the proportion that each category represents within the respective dataset. We normalised the job description percentages to 100% for easier comparison with the curriculum and university percentages, and the results are presented in Figure 5.

The mapping between higher education curricula and software engineering job advertisements reveals both strong alignment in foundational areas and notable quantitative discrepancies in others. These discrepancies range from moderate differences of approximately 10–20% to substantial gaps of nearly threefold, indicating that some skill areas are either underrepresented or overemphasised in UK university curricula relative to explicit industry demand.

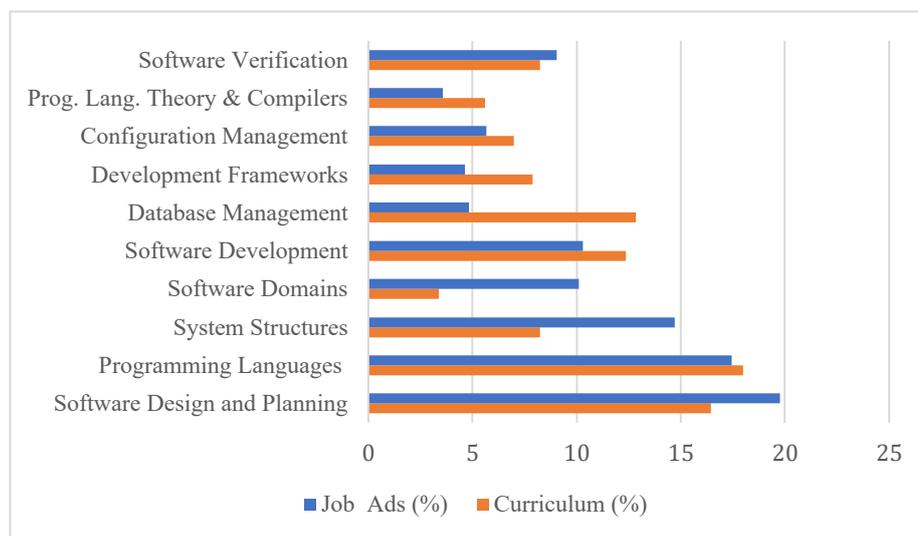

Figure 5. Frequency analysis of skills under each category

#### 4.3.1 Major Overemphasis in University Curricula

The results identify three skill categories where university curriculum emphasis substantially exceeds industry demand, as shown in Table 8.

*Database Management* shows the most pronounced overemphasis. Universities allocate 12.83% of curricular focus to this category, compared to only 4.83% in job advertisements. This corresponds to a ratio nearly three times higher in curricula than is explicitly requested by employers. This finding contradicts earlier international studies, such as those by Oguz et al. [2], which identified a lack of database skills in Turkish university curricula. The lower demand observed in UK job advertisements may suggest that database skills are often treated as implicit baseline knowledge in industry, particularly for backend or full-stack roles. Additionally, industry demand may increasingly favour the use of integrated databases within cloud and data engineering contexts, rather than standalone database theory, which remains prominent in academic programmes.

*Development Frameworks* also appear overemphasised, with nearly two times less frequently mentioned in job postings than in the curriculum. However, this discrepancy does not necessarily imply that frameworks are over-taught. Job advertisements often refer to broader concepts such as *"frontend development"* without listing specific frameworks (e.g., React or Angular), assuming familiarity with commonly used tools. Universities may therefore emphasise frameworks to provide tangible, practical skills, even if specific tools evolve rapidly in industry.

*Programming Language Theory and Compilers* has a 37% lower industry emphasis, which may reflect the more theoretical and academic nature of this category, primarily relevant for specialised roles and less frequently required in general software engineering positions.

Table 8. Alignment of higher education modules with job industry demand

| Category | Curriculum % | Job Ads % | Alignment | Comment |
|---|---|---|---|---|
| Software Design and Planning | 16.46 | 19.75 | Undervalued | Job demand is ~20% higher than curriculum coverage. |
| Programming Languages | 18.01 | 17.44 | Aligned | Near-perfect alignment (~18% vs. ~17.5%) |
| System Structures | 8.23 | 14.71 | Undervalued | A significant gap; job demand is ~79% higher than university coverage |
| Software Domains | 3.39 | 10.09 | Undervalued | Job demand is ~198% higher, indicating a major deficit in domain-specific knowledge |
| Software Development | 12.35 | 10.30 | Aligned | Curriculum coverage is ~17% higher, indicating a modest overemphasis on methodologies |
| Database Management | 12.83 | 4.83 | Overvalued | University emphasis is ~165% higher than explicit job ad demand |
| Development Frameworks | 7.89 | 4.62 | Overvalued | Curriculum coverage is ~71% higher than job ad mentions |
| Configuration Management | 7.00 | 5.67 | Overvalued | Curriculum coverage is ~23% higher |
| Programming Language Theory & Compilers | 5.62 | 3.57 | Overvalued | Academic focus is ~57% higher than industry demand |
| Software Verification & Validation | 8.24 | 9.04 | Aligned | Close alignment, with job demand ~10% higher |

### 4.3.2 Skill Gaps and Undervalued Areas in Curricula

The analysis also reveals several areas where industry demand significantly exceeds curricular coverage, indicating critical skill gaps.

*System Structures* represents the most prominent architectural gap. Job advertisements emphasise this category at 14.71%, while curricula allocate only 8.23%, meaning industry demand is approximately 79% higher than university coverage. This gap suggests that graduates may be underprepared in areas such as cloud-native architectures, distributed systems, scalability, and microservices. This finding aligns with prior research that also identified system-level knowledge as a major skills gap [10].

An even larger discrepancy is observed for *Software Domains*. Industry demand in this category reaches 10.09%, compared to only 3.39% in curricula, which is a three times gap, indicating that job advertisements frequently request domain-specific knowledge (e.g., healthcare IT, fintech systems, embedded software), whereas university programmes tend to prioritise general-purpose software engineering skills. This suggests that universities are primarily producing generalist graduates, while the labour market increasingly values specialised domain expertise.

*Software Design and Planning* also appears undervalued, though to a lesser extent. Job advertisements reference this category in 19.75% of cases, compared to 16.46% curriculum coverage, reflecting a 20% undervaluation gap. This suggests that industry places a slightly greater emphasis on high-level design, architectural planning, and systems thinking than is currently reflected in academic programmes.

### 4.3.3 Strong Alignment between Academia and Industry

Several skill categories demonstrate strong alignment between university curricula and job market demand. *General Programming Languages* shows near-perfect alignment, with a difference of only 3%. This suggests that foundational programming language instruction in UK universities closely matches industry expectations. *Software Verification and Validation* also exhibits close alignment. with 10% difference. This finding contrasts with studies by Sahin and Çelikkan [1] and Garousi et al. [4], which identified testing as a major skills gap, suggesting that UK curricula may have made progress in integrating verification and validation practices.

*Software Development* shows moderate alignment, with curricula emphasising this category at 12.35%, compared to 10.30% in job advertisements, a 17% difference. This may reflect universities' stronger focus on formal development methodologies (e.g., Agile, TDD, object-oriented principles), which are often assumed rather than explicitly stated in job postings.

### 4.4 Relationship between Location, Job Description, and Nature of the Role

The job market has several underlying correlation factors when examining the relationship between the type of job advertised, the nature of the role, and its location. The job adverts dataset enabled the analysis of the existing interdependency between the three areas.

To begin with, the nature of the role has seen an increasing number of jobs that were traditionally expected to be onsite being migrated towards remote options in recent years. For our dataset, the analysis results are presented in Table 9.

Table 9. Nature of the role distribution

| Nature | Count | Percentage |
|---|---|---|
| Onsite | 163 | 54.33% |
| Remote | 78 | 26.00% |
| Hybrid | 59 | 19.67% |

It is interesting to see that the "Covid effect" on jobs being typically remote is shrinking in favour of onsite positions, which appear to dominate the dataset. While the broader job market shows a strong post-pandemic shift back to onsite work, software engineering roles demonstrate a significant and sustained deviation from this trend. In our dataset, 26% of software engineering roles were advertised as fully remote and just under 20% of job advertisements allowed for a hybrid approach. This proportion is more than double the 9.5% observed in a study of AI engineering roles in the UK [15]. To summarise, the market is no longer "remote first," but rather remote-normalised. The job family also had a slight bias towards the nature of the job; we aimed to cluster the ads into two main clusters, with "engineer" and "developer" being the primary clusters (totalling 293 out of the 300 jobs), as shown in Table 10. As shown in the table, engineers are still predominantly onsite, reflecting the need for hardware interaction, combined with potentially regulated activities and lab-based work. In contrast, developers show the strongest propensity for remote work, with almost half of the remote roles (43%) being developer positions. Finally, hybrid positions provide a compromise solution for engineering roles. Beyond the tabular data, we examined the statistical relationship between the nature of the job and the job family for the two job clusters (developers and engineers). A chi-square test of independence yielded $\chi^2(2) = 6.04$ and $p = 0.0488$. At $\alpha = 0.05$, we reject the null hypothesis of independence, indicating that job nature is statistically associated with job family.

Table 10. Relationship between job title and nature of the role

| Job Family | Onsite | Remote | Hybrid |
|---|---|---|---|
| Engineer | 110 | 41 | 37 |
| Developer | 48 | 36 | 21 |

Moving to the relationship between the city and the nature of the job, the analysis is significantly skewed by London, which accounted for just under a quarter of the adverts. More significantly, remote jobs were typically not tied to a specific city. Out of the 73 job adverts listed for London, there were only 13 hybrid jobs and just one remote listing. At the other end of the spectrum, "unknown" locations had 46 out of 56 jobs remote, nine hybrid, and, possibly due to an elliptical advert, one onsite.

We repeated the chi-square analysis to determine whether the nature of the job is independent of the city in which it is advertised. To avoid distortion from small sample sizes, we included only cities with at least five roles (11 cities in total, covering the majority of postings) and excluded advertisements with an unspecified city. The test yielded $\chi^2(20) = 182.12$, $p = 3.7 \times 10^{-28}$. This result is extremely statistically significant, providing very strong evidence to reject the null hypothesis of independence and indicating that the distribution of onsite, hybrid, and remote roles varies significantly by city. Overall, job location shows a much stronger association with job nature than job title.

#### 4.5 Limitations

Ten categories were established, and skills from both university modules and job descriptions were successfully assigned to these categories. While this structured approach enables comparison across educational and industry datasets, it is subject to inherent limitations. The categorisation process involves subjective decisions regarding which terms and modules belong to each category, which can introduce inconsistencies in the interpretation and valuation of skills. The relatively small number of categories further compounds this issue, as it may oversimplify the diversity of skills into broad groupings, potentially overlooking important competencies that do not fit neatly within the predefined schema.

In addition, the reliance on keyword-based fuzzy matching introduces further constraints. Although this method enables automated and scalable skill identification, it depends on predefined keyword lists and string similarity, limiting its ability to capture contextual meaning, semantic variation, and newly emerging skill expressions. As a result, this limitation can affect the interpretation of identified skill gaps, especially for evolving technologies that are not yet explicitly represented within established frameworks such as ACM CCS or SWEBOK, including cloud-native frameworks and AI pipeline-related skills. The exclusion of ambiguous or short terms to reduce false positives, while necessary for improving precision, may also contribute to the underrepresentation of certain competencies.

A significant example of this limitation is the exclusion of emerging technologies such as cloud computing and blockchain from the ACM CCS taxonomy, which underpins the categorisation framework. Given the growing importance of these technologies in the digital economy, their absence highlights the constraints of the current taxonomic framework and underscores the necessity for ongoing revision and updating of categorisation schemes to incorporate emerging technologies.

## 5  Discussion

#### 5.1 Skill Gaps Patterns

The results reveal a clear pattern: the largest gaps occur where practical, industry-specific knowledge is required. The most substantial gaps appear in *Software Domains*, which has a three-times gap, and System Structures, which has nearly a two-times gap between industry demand and skills offered by the universities. This suggests that industry demand for competencies related to system architectures, distributed systems, cloud computing, and microservices is substantially higher than the emphasis placed on these topics in university programmes. This finding aligns with prior research [10] and reflects the growing architectural complexity of modern software systems.

A gap in *Software Domains* indicates that while universities tend to produce broadly trained graduates, employers increasingly seek candidates with contextual expertise in specific application domains such as fintech, healthcare systems, or embedded software. The scale of this difference highlights the challenge universities face in balancing generalist education with industry specialisation.

These gaps, alongside the overemphasis of curriculum on *Database Management and Development Frameworks,* suggest that universities prioritise theoretical foundations over practical, industry-relevant architecture and domain knowledge.

Database-related skills appear in curricula at nearly three times the rate at which they are explicitly requested by employers. Several factors may explain this pattern. First, database knowledge is often treated as implicit baseline competence for backend or full-stack roles and, therefore, may not be explicitly stated in job descriptions. Second, keyword-based analysis may undercount database demand when job advertisements refer to broader terms, such as "backend development," without specifically naming SQL or specific database technologies. Finally, database courses have long been a core component of computer science curricula**,** and universities may continue to prioritise standalone database theory, while industry increasingly expects these skills to be integrated with cloud platforms and data engineering practices.

A similar situation with *Development Frameworks* subjects, with university coverage almost twice as high as job advertisement mentions. This difference does not necessarily imply that frameworks are taught excessively. Rather, employers may assume familiarity with common frameworks such as React or Angular and therefore omit them from listings, or refer instead to broader competencies like "frontend development". From an educational perspective, frameworks serve as concrete entry points for applying programming knowledge, even if the specific tools evolve rapidly in industry.

Assuming that the industry considers some skills as baseline competencies can lead to their frequent exclusion in job descriptions, regardless of curricular coverage. Additionally, the rapid pace of technological adoption in the workplace means that new frameworks, tools, and methodologies can emerge faster than they are incorporated into academic programs. While these factors may help explain why some gaps persist in even well-established computer science and software engineering courses, an analysis of the underlying causes of skill gaps is beyond the scope of this chapter and has been addressed in an earlier study by Diniz et al. [8]. The authors provide a detailed analysis, identifying causes such as misalignment between academia and industry, limited practical experience of teaching staff, unstructured curricula, and insufficient exposure to real-world projects.

**5.2 Recommendations**

To address this imbalance in curriculum, enhance graduate employability and align with explicit market needs, a strategic reallocation of curricular focus is necessary. We recommend reducing the proportion of standalone, theoretical database instruction and integrating its core principles within applied modules on cloud data engineering and system design. Concurrently, dedicated modules on Modern Software Architecture, covering cloud-native patterns, microservices, and distributed systems, should be introduced or expanded to enhance architectural education. To bridge the significant gap in domain-specific knowledge, institutions should develop sector-focused elective pathways (e.g., FinTech, Health Informatics) in partnership with industry, ensuring theoretical learning is contextualised within real-world applications.

Furthermore, strengthening the bridge between academia and industry is crucial. Establishing formal Industry Advisory Boards for periodic curriculum review. Inviting industry representatives as guest lecturers and running industry-sponsored final projects will ensure educational content remains responsive to evolving technological and sectoral demands. Implementing these evidence-based changes will enable higher education to produce graduates equipped with a relevant, applied skill set that matches the explicit needs of the software engineering job market.

As highlighted by prior studies, significant pedagogical challenges remain even within a well-aligned curriculum. Students may struggle to grasp the proper understanding of complex subjects, such as abstract system architectures or concurrent programming models [28]. Furthermore, the uneven distribution of focus and effort in group-based project work, a staple of software engineering education, can lead to inconsistent skill acquisition among graduates [29]. Therefore, while our recommendations focus on recalibrating curricular content to meet industry demands, parallel attention must be paid to pedagogical delivery and assessment design. Addressing these intrinsic learning challenges is essential to ensure that the potential of an aligned curriculum is fully realised, enabling all students to effectively integrate and apply the targeted knowledge and skills.

Furthermore, these findings provide practical insights for UK policy and accreditation processes by highlighting which technical and domain-specific competencies should be prioritised in computing and software engineering programs, thereby supporting the alignment of curricula with evolving industry requirements.

A comparison with existing research further illustrates the contextual nature of our findings (see Table 11). While database management appears overemphasised in the UK, it was reported as undervalued in Turkish universities by Oguz et al. [2]. Conversely, the prominence of System Structures as an area of misalignment is consistent with earlier work [10], suggesting this may be a more persistent and widespread issue. Discrepancies regarding Software Design and Testing across studies may reflect differences in how these categories are defined, the time period of analysis, or the specific industry sectors represented in the data.

Table 11. Comparative analysis with existing literature

| Finding | Our Study | Previous Studies | Comments |
| --- | --- | --- | --- |
| Database Management | Overemphasised | Undervalued [2] | Turkish vs. UK context |
| System Structures | Skill gap | Supported [10] | Consistent finding |
| Software Design | Not prominent gap | Skill gap [2], [4] | Discrepancy in findings |
| Testing/Validation | Aligned | Skill gap [1], [4] | Regional/cultural differences |

While this chapter focuses on technical skills, soft skills are widely recognised as critical to employability in software engineering. A study by Matturro et al. [30] conducted a systematic literature review to identify relevant soft skills in software engineering, reviewing 44 papers and identifying 30 main categories. They found that communication (91%), teamwork (68%), analytical skills (55%), and organisational/planning skills (55%) were the most frequently mentioned. Other important skills include leadership, problem-solving, autonomy, and initiative. Graduates lacking these abilities may struggle to work effectively in teams, interact with clients, or manage complex projects.

# 6   Conclusions

This chapter aimed to answer the research question: *What are the prominent technological knowledge gaps between higher education institutions and the Software Engineering industry in the UK?* Through an automated analysis mapping skill category frequencies in module curricula against job advertisements, we have identified and quantified several critical misalignments.

Our findings clearly indicate that the most prominent gaps are areas of applied and integrative knowledge. Specifically, the categories System Structures, which cover modern architectures such as cloud and distributed systems, and Software Domains, encompassing sector-specific contexts like FinTech or Health Informatics, are significantly underrepresented in HEI curricula compared to industry demand. Conversely, the category Database Management receives considerably more emphasis in academic programmes than is explicitly sought in job market descriptions.

These results underscore a fundamental insight: the technological skills landscape evolves at a pace that challenges traditional, static curriculum design. Manual methods for gap analysis are often too slow to capture these shifts. Therefore, the automated, data-driven methodology developed and employed in this project is not merely an academic exercise but a necessary tool. It provides a viable mechanism for HEIs to continuously monitor, reassess, and dynamically adapt their computer science and software engineering curricula. By adopting this approach, institutions can ensure that their graduates are equipped with the relevant, timely technological knowledge required to thrive in their careers, thereby directly addressing the identified gaps.

### 6.1 Future Work

For future work, the data foundation could be strengthened by expanding the web scraper to include a wider range of job boards, capturing a more comprehensive view of the market. Second, a significant limitation was the focus solely on technical skills; integrating an analysis of soft skills, which are frequently cited in the literature as crucial, would offer a more comprehensive picture of graduate readiness [30, 31].

Most importantly, the accuracy of the skill-matching engine could be substantially improved. Moving beyond fuzzy matching to employ Named Entity Recognition (NER) models would allow the system to understand context and synonyms [32]. For instance, a custom NER model could disambiguate "Java", the programming language, from "Java", the island, and identify skill phrases expressed in various languages. This advancement would reduce false positives and provide a deeper, more reliable analysis of evolving skill trends, further automating the critical task of curriculum gap analysis.